\documentstyle[12pt,epsfig]{article}
\newcommand{\npb}[3]{Nucl.~Phys.~B #1 (#2) #3}
\newcommand{\plb}[3]{Phys.~Lett.~B #1 (#2) #3}

\newcommand{\cpc}[3]{Comput.~Phys.~Commun. #1 (#2) #3}
\begin{document}
\bibliographystyle{unsrt}

\rightline{FNT/T-2001/13}
\begin{center}
{\Large\bf Quartic anomalous couplings at LEP}
\end{center}
\vskip 24pt
\begin{center}
{\large
G.~Montagna$^a$, M.~Moretti$^b$, O.~Nicrosini$^c$, 
M.~Osmo$^a$ and F.~Piccinini$^c$}
\end{center}
\vskip 24pt
\begin{center}
$^a$ Dipartimento di Fisica Nucleare e Teorica 
- Universit\`a 
di Pavia, and \\
INFN - Sezione di Pavia, Via A. Bassi 6, Pavia, Italy\\
\vskip 12pt\noindent
$^b$ Dipartimento di Fisica - Universit\`a di    Ferrara, and \\
INFN - Sezione di Ferrara, Ferrara, Italy
\vskip 12pt\noindent
$^c$ INFN - Sezione di Pavia, and\\ 
Dipartimento di Fisica Nucleare e Teorica 
- Universit\`a 
di Pavia, \\ Via A. Bassi 6, Pavia, Italy\\
\end{center}
\vskip 48pt
\begin{abstract}
The search for quartic anomalous gauge couplings at LEP requires 
appropriate predictions for the radiative processes 
$e^+ e^- \to \nu\bar\nu \gamma\gamma$, $e^+ e^- \to q\bar{q}\gamma\gamma$ and 
$e^+ e^- \to 4$~fermions+$\gamma$. Matrix elements are exactly computed 
at the tree level, and the
effects of anomalous couplings and initial-state radiation are included. 
Comparisons with results and approximations existing in the literature
 are shown and commented.
Improved versions of the event generators {\tt NUNUGPV} and 
{\tt WRAP}  are
made available for experimental analysis.

\end{abstract}
\newpage

\section{Introduction}
\label{intro}
Despite of the striking success of the Standard Model (SM) in accommodating
the precision data collected at high-energy colliders, important 
tests of the theory, such as the non-abelian nature of the gauge 
symmetry and the mechanism of electroweak symmetry breaking, are still
at a beginning stage. To this end, gauge-boson self interactions 
play a key role. Presently, triple gauge couplings are being 
probed at LEP~\cite{leptri} and Tevatron~\cite{tevatron}, while only very recently direct 
measurements of quartic couplings became available through 
the study of radiative events 
at LEP~\cite{l3quad}-\cite{mm}. Actually, 
events with one or two isolated, hard photons are analyzed at LEP 
to search for anomalies in the sector of quartic gauge-boson couplings.
Only vertices involving at least one photon can be constrained since 
quadrilinear interactions containing four massive gauge bosons give 
rise to a three massive gauge boson final state and 
are therefore beyond the potentials
of LEP due to lack of phase space. The processes considered in the 
experimental analyses are $e^+ e^- \to W^+ W^- \gamma$, 
$e^+ e^- \to Z \gamma \gamma$ and 
$e^+ e^- \to  \nu \bar\nu \gamma \gamma$~\cite{l3quad}-\cite{mm}.
The $W^+ W^- \gamma$ signature, which yields a four fermion plus gamma 
final state, is interesting in order to test $WW\gamma \gamma$ 
and $WWZ\gamma$ vertices. For the $Z \gamma \gamma$ events, the 
final states due to the hadronic decays of the $Z$ boson with two 
jets and two visible photons are selected to probe the purely anomalous
vertex $ZZ\gamma \gamma$, which is of particular interest being absent in the
SM at tree level. The final state with two neutrinos and 
two acoplanar photons allows to study the quartic $WW\gamma\gamma$ 
and $ZZ\gamma\gamma$ interactions. 

For tests of the SM and searches for new physics beyond it, anomalous 
quartic gauge-boson couplings are important because, as widely discussed 
in the literature~\cite{book}, they offer a window on the mechanism of symmetry breaking 
and it is possible to imagine extensions of the SM that alter the quartic
vertices without modifying the trilinear interactions. For this reason, 
phenomenological analyses of anomalous quartic interactions have already 
appeared in the literature, by considering the potentials of $\gamma \gamma$~\cite{bb}, 
$e \gamma$~\cite{an} and $e^+ e^-$~\cite{an1}-\cite{bela} colliders and very recently also of 
hadron machines~\cite{qagch}. 
The tightest constraints on quartic anomalous couplings come from 
electroweak precision data, as discussed 
in ref.~\cite{qagch}. However, present measurements at LEP2 can investigate photonic quartic couplings
for the first time in a direct way through the study of the final states 
above discussed. In the light of these experimental analyses, the aim of 
this letter is to present exact SM calculations of the processes 
$e^+ e^- \to 4$~fermions ($4f$) + $\gamma$, 
$ e^+ e^- \to q \bar{q} \gamma \gamma$ and 
$e^+ e^- \to  \nu \bar\nu \gamma \gamma$, including the effects of quartic
anomalous gauge couplings (QAGC) and of the most important radiative corrections
due to initial-state radiation (ISR). The sensitivity of LEP searches is 
studied in comparison with approximate results existing in the 
literature. It turns out that the $Z\gamma\gamma$ approximation works 
well for the $q \bar{q} \gamma \gamma$ and $\nu \bar\nu \gamma \gamma$
 channels when appropriate cuts around the $Z$ mass are imposed, 
 while the $WW\gamma$ approximation can significantly differ from the
 exact $4f$ + $\gamma$ calculation, even in the presence of 
 invariant mass cuts around the $W$ mass. It is also shown that 
 the $\nu \bar\nu \gamma \gamma$ final state with appropriate cuts on the
 recoil mass can be successfully exploited to extract limits on neutral QAGC, 
 as a complementary channel to the $q \bar{q} \gamma \gamma$ process.

\section{Theoretical approach}
\label{sec:1}
The theoretical framework of interest is the formalism of electroweak 
chiral lagrangians. In such a scenario, QAGC involving four massive 
gauge bosons emerge as operators of dimension four at 
next-to-leading order,  
while QAGC with at least one photon originate from 
six (or higher) dimensional operators at next-to-next-to-leading order. 
They are said genuinely anomalous if they do not induce 
new trilinear gauge interactions. 

Anomalous $WW\gamma\gamma$ and $ZZ\gamma\gamma$ vertices 
were originally introduced in ref~\cite{bb}. In this paper, the authors 
show that, by assuming $C$ and $P$ conservation and 
further imposing $U(1)_{em}$ gauge invariance and $SU(2)_c$ custodial 
symmetry,  two independent Lorentz structures 
contribute to $WW\gamma\gamma$ and $ZZ\gamma\gamma$ interactions 
according to the following lagrangians 
\begin{eqnarray}
{\cal L}_{0} &=& - \frac{e^2}{16} \, { \frac{a_0}{\Lambda^2} } \,
F_{\mu\nu}F^{\mu\nu} \vec{W}^\alpha \cdot {\vec W}_{\alpha} \nonumber\\
{\cal L}_{c} &=& - \frac{e^2}{16} \, { \frac{a_c}{\Lambda^2} } \,
F_{\mu\alpha}F^{\mu\beta} \vec{W}^\alpha \cdot {\vec W}_{\beta} \, ,
\label{bb}
\end{eqnarray}
where $F_{\mu\nu} = \partial_\mu A_\nu - \partial_\nu A_\mu$ is the 
electromagnetic field tensor, and $\vec{W}$ is a $SU(2)$ triplet describing 
the $W$ and $Z$ physical fields, i.e. 
\begin{eqnarray}
\vec{W} = \left ( 
            \begin{array}{c}
	    \frac{1}{\sqrt{2}}(W_\mu^+ + W_\mu^-) \\
	    \frac{i}{\sqrt{2}}(W_\mu^+ - W_\mu^-) \\
	    Z_\mu/\cos\theta_w
	    \end{array}
	    \right ) \nonumber \, ,
\end{eqnarray}
$\cos\theta_w$ being the cosine of the weak mixing angle. In eq.~(\ref{bb}) 
$a_0$ and $a_c$ are (dimensionless) anomalous couplings, divided 
by an energy scale $\Lambda$, which has the meaning of scale of new physics.
Generally speaking, $\Lambda$ is in principle unknown and model-dependent. 
However, the ratios $a_i/\Lambda^2$ entering the phenomenological lagrangians
can be meaningfully extracted from the data in a model-independent way.

An anomalous $WWZ\gamma$ interaction has also been advocated 
in refs.~\cite{an}-\cite{swer}. The 
corresponding phenomenological lagrangian reads as follows
\begin{eqnarray}
{\cal L}_{n} = - \frac{e^2}{16} {\frac{a_n}{\Lambda^2} }
\epsilon_{ijk} W_{\mu\alpha}^i W_{\nu}^j W^{k\alpha} F^{\mu\nu} \, ,
\label{an}
\end{eqnarray}
where $\vec{W}_{\mu\nu}$  is the $SU(2)$ field strength tensor, and 
$a_n/\Lambda^2$ the anomalous coupling. Although 
$U(1)_{em}$ and $SU(2)_c$ conserving, the lagrangian of eq.~(\ref{an}) 
violates $C$ and $CP$, as noticed in the literature~\cite{bela}. 
The anomalous couplings $a_0,a_c,a_n$ entering eqs.~(\ref{bb})-(\ref{an}) 
are presently constrained by LEP 
collaborations~\cite{l3quad}-\cite{mm}. 
Under the assumptions 
of local $U(1)_{em}$ invariance and global custodial $SU(2)_c$ symmetry,
two additional operator structures which violate $P$ have been very recently 
proposed in ref.~\cite{ddrw}. 

A complete and general analysis of photonic quartic couplings
has been performed in ref.~\cite{bela}. In this paper it is shown that, by imposing $C$, 
$P$ and $U(1)_{em}$ invariance, nine independent Lorentz structures do 
contribute to $WW\gamma\gamma$, $ZZ\gamma\gamma$ and $WWZ\gamma$ vertices. It is 
further demonstrated by the authors that, by embedding these structures in $SU(2)\times U(1)$ 
gauge invariant and $SU(2)_c$ symmetric combinations, fourteen $C$ and $P$ conserving
operators are allowed, with $k_j^i$ parameters, which parameterize the strength 
of anomalous couplings. 

Following ref.~\cite{bela}, the nine independent operator structures 
contributing to the vertices analyzed at LEP have been implemented 
directly at the lagrangian level in the 
{\tt ALPHA} algorithm~\cite{alpha}, according to 
the formula 
\begin{eqnarray}
L_{QAGC} =&& W_1 + W_2 + Z_1 +  Z_2 \nonumber\\
   && W_0^Z +  W_c^Z + W_1^Z + 
    W_2^Z +  W_3^Z \, ,  
\label{wwzgan1}
\end{eqnarray}
where the Lorentz structure of the operators is given by:
\begin{equation}
\begin{array}{llllll}
W_1  = & a_{w1} F_{\mu\nu}F^{\mu\nu}W^+_{\rho}W^{-\rho} \\
W_2 = & a_{w2} F_{\mu\nu}F^{\mu\rho}W^{+\nu}W^-_{\rho}+ {\rm h.c.}\\   
Z_1  = & a_{z1} F_{\mu\nu}F^{\mu\nu}Z_{\rho}Z^{\rho} \\
Z_2 = & a_{z2} F_{\mu\nu}F^{\mu\rho}Z^{\nu}Z_{\rho}\\   
W_0^{Z} = & a_{wz0} F_{\mu\nu}Z^{\mu\nu}W^+_{\rho}W^{-\rho} \\ 
W_c^{Z} = & a_{wzc} F_{\mu\nu}Z^{\mu\rho}W^{+\nu}W^-_{\rho}+ {\rm h.c.}\\
W_1^{Z} = & a_{wz1} F_{\mu\nu}W^{+\mu\nu}Z^{\rho}W^-_{\rho}+ {\rm h.c.}\\
W_2^{Z} = & a_{wz2} F_{\mu\nu}W^{+\mu\rho}Z^{\nu}W^-_{\rho}+ 
{\rm h.c.}\\ 
W_3^{Z} = & a_{wz3} F_{\mu\nu}W^{+\mu\rho}Z_{\rho}W^{-\nu}+ {\rm h.c. } \, ,
                             && 
			     \end{array}
\label{wwzgan2}
\end{equation}
the $a_i$ being coefficients of dimension $M^{-2}$.

The first two structures refer to $WW\gamma\gamma$ interactions, the third and 
fourth ones to $ZZ\gamma\gamma$ interactions, while the remaining five affect
the $WWZ\gamma$ vertex. It is worth 
noticing that, by virtue of eq.~(\ref{wwzgan2}), both parameterizations available 
in the literature for QAGC, namely the parameterization in terms of $a_0, a_c, a_n$ 
couplings and the one in terms of $k_j^i$ coefficients, can be obtained 
by means of appropriate relations between the $a_i$ parameters. For example, 
 assigned $a_0$, $a_c$ and $a_n$, the following relations for 
 the $a_i$
coefficients of eq.~(\ref{wwzgan2}) hold
\begin{equation}
\begin{array}{lcl}
a_{w1} = -\frac{e^2}{8 \Lambda^2} \, a_0\\
a_{z1} = -\frac{e^2}{16 \cos^2\theta_w \Lambda^2} \, a_0\\
a_{w2} = -\frac{e^2}{8 \Lambda^2} \, a_c\\
a_{z2} = -\frac{e^2}{16 \cos^2\theta_w \Lambda^2} \, a_c\\
a_{wzc} = i \frac{e^2}{16 \cos\theta_w  \Lambda^2} \, a_n \\
a_{wz2} = i \frac{e^2}{16 \cos\theta_w  \Lambda^2} \, a_n  \\
a_{wz3} = -i \frac{e^2}{16 \cos\theta_w  \Lambda^2} \, a_n \, .
\end{array}
\end{equation}

The implementation in {\tt ALPHA} has been 
carefully cross-checked by an analytical 
calculation of the scattering processes $WW \to \gamma\gamma$, 
$ZZ \to \gamma\gamma$ and $WW \to Z\gamma$ (and its permutations), 
finding perfect agreement. 

In order to provide valuable tools for experimental data 
analysis, the version of {\tt ALPHA} including the lagrangian 
of eq.~(\ref{wwzgan2}) has been interfaced to the Monte Carlo
generators {\tt NUNUGPV}~\cite{nunu} and {\tt WRAP}~\cite{wrap}, 
for a complete study of QAGC in radiative events at LEP. 
The $\nu\bar\nu\gamma\gamma$ final state can be simulated by means 
of  {\tt NUNUGPV}, which includes exact SM matrix elements and 
$p_t$-dependent QED Structure Functions (SF) to account for ISR. 
The other processes of interest at LEP 
can be studied by using {\tt WRAP}, which is based on the exact matrix 
elements for the (charged-current) reactions $e^+ e^- \to 4f$+$\gamma$ and 
the process $e^+ e^- \to q \bar{q} \gamma\gamma$ and the same 
treatment as  {\tt NUNUGPV} for ISR. By means of {\tt WRAP}, predictions
for the inclusive final states $WW\gamma$ and $Z\gamma\gamma$ can be also
obtained, especially in order to compare with results existing in the 
literature treating the $W$ and $Z$ bosons in the on-shell approximation, 
as those of refs.~\cite{swer,bela}.

\begin{figure*}
\begin{center}
\epsfig{file=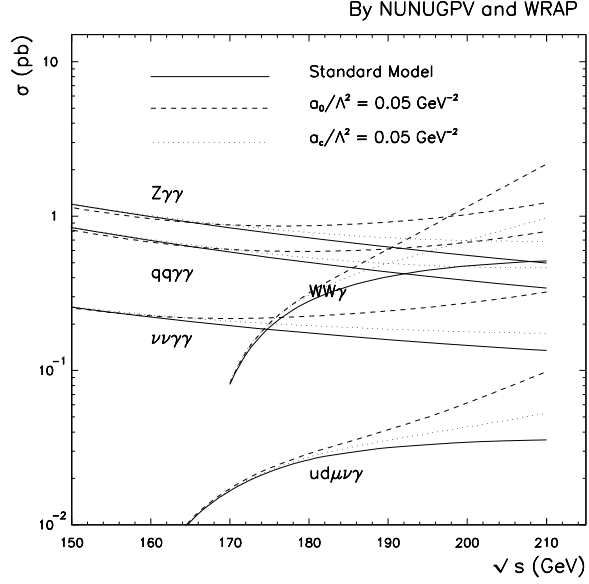,height=9.truecm}
\caption{\small The cross sections in the absence (solid line) 
and in the presence (dashed and dotted lines) 
of anomalous parameters $a_0$ and $a_c$ 
($a_0/\Lambda^2 = a_c/\Lambda^2$ = 0.05~${\rm GeV}^{-2}$) for the processes 
$e^+ e^- \to Z\gamma\gamma, WW\gamma, q \bar{q} \gamma\gamma,  \nu\bar\nu \gamma\gamma, 
u \bar{d} \mu \bar{\nu}_{\mu}\gamma$, as functions of the c.m. energy in the 
LEP2 energy range.}
\label{enscan}
\end{center}
\end{figure*}

\section{Numerical results and discussion}

In order to test the implementation of QAGC, comparisons with 
the available results have been performed. A first comparison has 
been done with the plots published in ref.~\cite{bela} for the dependence of the 
$WW\gamma$ and $Z\gamma\gamma$ cross sections on the anomalous 
parameters $k_i^j$ at $\sqrt{s}$ = 200 GeV, using the same 
set of input parameters and cuts. The results of such a comparison 
between the results of ref.~\cite{bela} 
and the predictions of the Monte Carlo {\tt WRAP}
show a very satisfactory agreement, testifying a correct implementation 
of $WW\gamma\gamma$, $ZZ\gamma\gamma$ and $WWZ\gamma$ couplings 
in the present calculation.  

As far as the parameterization in terms of $a_0$ and $a_c$ anomalous 
couplings is concerned, two 
further comparisons have been performed. The first one has been 
done with the results of ref.~\cite{bela} for the 
dependence of the $WW\gamma$ cross section on $a_0$, 
finding perfect agreement. The second one has been 
performed with the results published 
in ref.~\cite{swer}
for the dependence of the $WW\gamma$ and $Z\gamma\gamma$ cross sections
on $a_0$ and $a_c$ parameters, using the same set of cuts. 
The predictions obtained with {\tt WRAP} 
differ from those of ref.~\cite{swer}. More precisely, an opposite sign is present for the
relative effects of $a_0$ and $a_c$ on the $WW\gamma$ cross section, as 
noticed in ref.~\cite{ddrw} and confirmed~\cite{aw} by the authors of 
ref.~\cite{swer}, whereas  this is 
not the case for the $Z\gamma\gamma$ final state. Work is in progress in order 
to clarify this point~\cite{aw}.

\begin{figure*}
\begin{center}
\epsfig{file=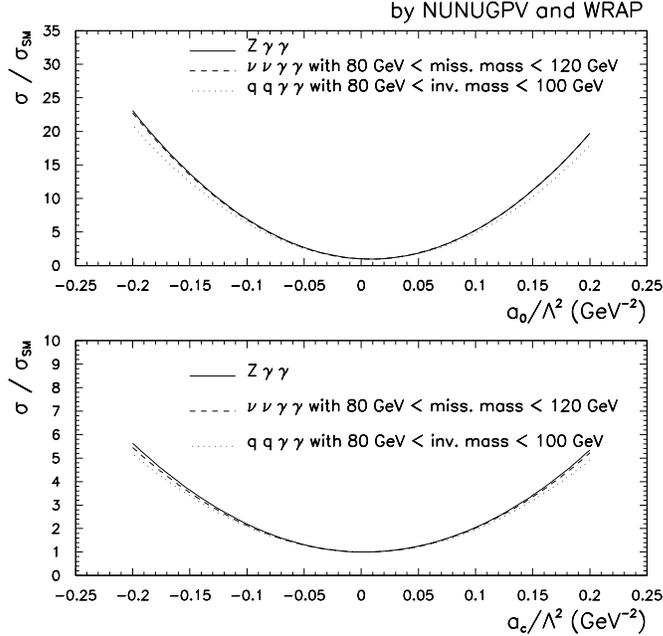,height=10.truecm}
\caption{\small The sensitivity of the 
$q \bar{q} \gamma\gamma$ and $\nu\bar\nu \gamma\gamma$ final states to 
$a_0$ and $a_c$ anomalous 
couplings in comparison with the $Z\gamma\gamma$ approximation, 
at $\sqrt{s} = $200~GeV.}
\label{a0acnngg}
\end{center}
\end{figure*}

For the results shown in the following, the input parameters used are:

\begin{equation}
\begin{array}{llll}
G_F=1.16637\cdot10^{-5}\hspace{2pt}{\rm GeV}^{-2} \hspace{.5cm} & 
M_Z=91.1867\hspace{2pt} {\rm GeV} \\
M_W=80.35 \hspace{2pt}{\rm GeV}
& \sin^2\theta_w = 1 - M_W^2/M_Z^2\\
\Gamma_Z=2.49471 \hspace{2pt} {\rm GeV} & \Gamma_W=2.04277 \hspace{2pt} 
{\rm GeV} \\
\end{array} 
\label{param}
\end{equation}
The cuts adopted are: $E_\gamma \geq 5$~GeV, $15^\circ \leq \vartheta_\gamma \leq 
165^\circ$ as detection criteria for the observed photons, 
together with a photon-final state charged fermion separation cut of $5^\circ$. 
Invariant mass cuts of the kind 80~GeV $\leq M_{q\bar{q}} \leq 100$~GeV for 
the $e^+ e^- \to q \bar{q} \gamma\gamma$ process and 
$M_{u\bar{d}} \geq$10~GeV for $e^+ e^- \to u \bar{d} \mu \bar{\nu}_{\mu}\gamma$
process are also imposed.

Figure \ref{enscan} shows the total cross sections, as 
obtained by means of the event generators {\tt NUNUGPV} and 
{\tt WRAP}, of all the radiative 
processes accessible to the LEP investigation, as functions of the 
center of mass (c.m.) energy in the LEP2 energy range. The solid lines 
correspond to the 
pure SM predictions; the dashed and dotted lines illustrate the effects of 
anomalous couplings 
$a_0/\Lambda^2 = 0.05~{\rm GeV}^{-2}$ and 
$a_c/\Lambda^2 = 0.05~{\rm GeV}^{-2}$, 
respectively. The unitarity violations induced by the anomalous couplings 
are clearly visible, their effects growing as the c.m. energy increases.

\begin{figure*}
\begin{center}
\epsfig{file=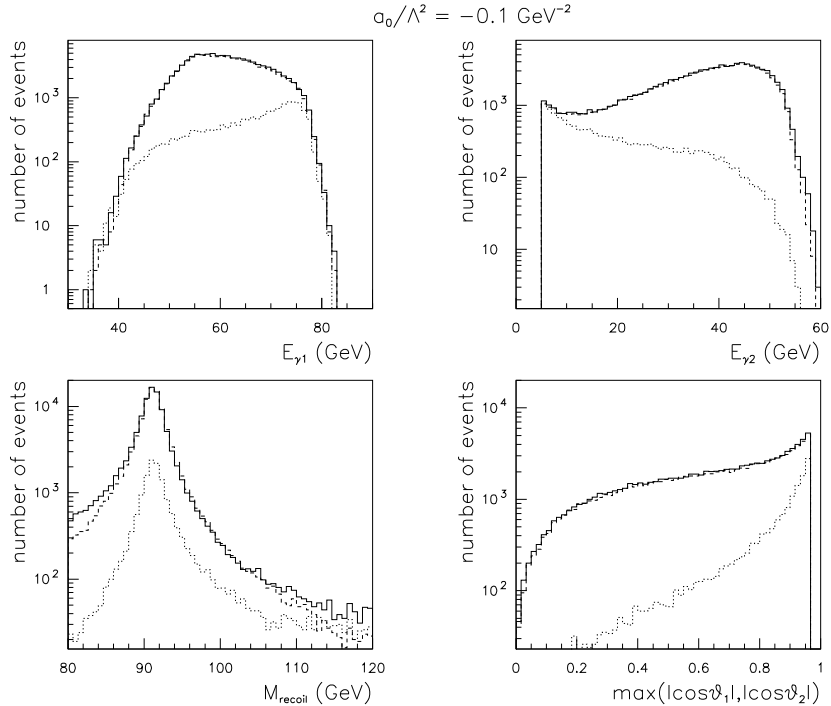,height=10.truecm}
\caption{\small Distribution of the energy of the most energetic photon
($E_{\gamma 1}$), the energy of the second most energetic 
photon ($E_{\gamma 2}$), the recoil mass 
($M_{\rm recoil}$) and the polar angle 
of the most forward photon (${\rm max}(|\cos\vartheta_1|,|\cos\vartheta_2|)$), 
for the $\nu\bar\nu \gamma\gamma$ final state, 
with $a_0/\Lambda^2 = -0.1~{\rm GeV}^{-2}$ 
and $\sqrt{s} = $200~GeV.
Dotted histograms: SM calculation; 
solid histograms: full $\nu\bar\nu \gamma\gamma$ calculation 
for $a_0/\Lambda^2 = -0.1~{\rm GeV}^{-2}$; dashed histograms: 
$\nu_{\mu}\bar\nu_{\mu} \gamma\gamma$ prediction 
for $a_0/\Lambda^2 = -0.1~{\rm GeV}^{-2}$.}
\label{a0d}
\end{center}
\end{figure*}

In Fig. \ref{a0acnngg} the sensitivity of the final states 
$q \bar{q} \gamma\gamma$ 
(dotted line) and 
$\nu\bar\nu \gamma\gamma$ (dashed line), as obtained by means 
of the full calculation, to the 
$a_0$ and $a_c$ parameters is shown at the c.m. energy of 200 GeV, in comparison with
the $Z\gamma\gamma$ approximation (solid line). 
The predictions for the $q \bar{q} \gamma\gamma$ channel 
refer to a cut on the invariant mass of the jet-jet system around the $Z$ mass 
(80~GeV $\leq M_{q\bar{q}} \leq 100$~GeV), while for the 
$\nu\bar\nu \gamma\gamma$ final state a cut on the recoil mass again around the 
$Z$ mass (80~GeV $\leq M_{recoil} \leq 120$~GeV) has been imposed. 
These cuts, also 
typically adopted in the experimental analysis,  
are required in order to compare with the $Z\gamma\gamma$ approximation, which, 
according to the results available in the literature~\cite{swer,bela}, 
assumes the $Z$ boson as an
on-shell particle. Therefore, this comparison allows to quantify the effects due to
the $\gamma$-$Z$ interference in the $q \bar{q} \gamma\gamma$ channel and to 
$W$-$Z$ interference in the $\nu\bar\nu \gamma\gamma$ one, as well as 
the effects of the off-shellness of the $Z$ boson, in the extraction of limits 
on the QAGC. It can be seen from Fig. \ref{a0acnngg} that the $Z\gamma\gamma$ 
calculation is a very good approximation of the full prediction, which includes
interference contributions. A detailed numerical investigation shows an 
agreement between the integrated cross sections 
at the per cent level as a function of $a_0$ and $a_c$ variations inside 
the presently allowed experimental constraints, thus 
illustrating the reliability of the $Z\gamma\gamma$ approximation in view of the 
expected experimental precision. 

\begin{figure*}
\begin{center}
\epsfig{file=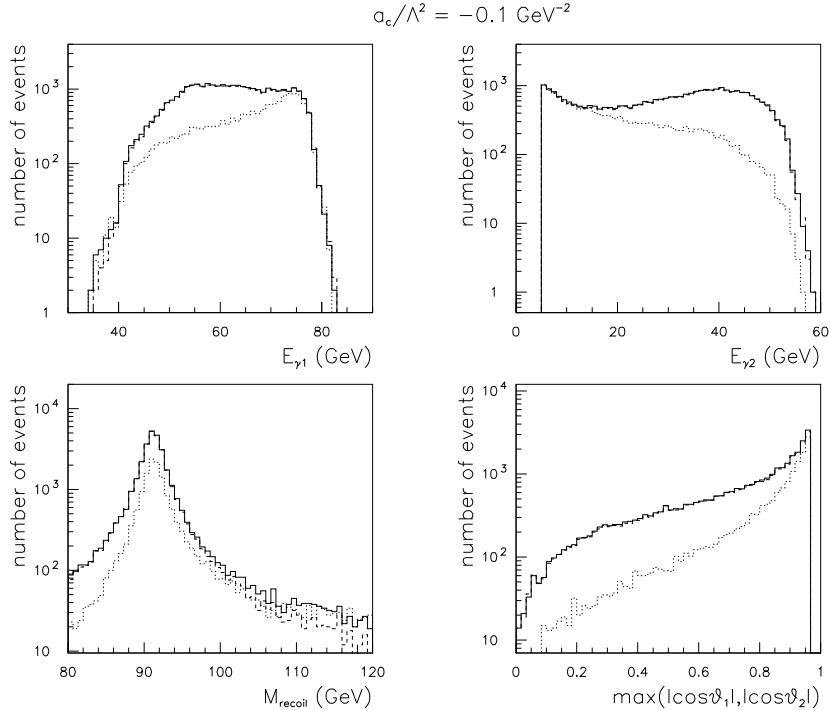,height=10.truecm}
\caption{\small The same as Fig.~\ref{a0d}, for 
$a_c/\Lambda^2 = -0.1~{\rm GeV}^{-2}$.}
\label{acd}
\end{center}
\end{figure*}

This conclusion also holds for the 
differential distributions mostly sensitive 
to QAGC and considered in the experimental studies, 
as can be noticed from Figs.~\ref{a0d}-\ref{acd}, showing the 
energy of the most energetic photon ($E_{\gamma 1}$), the energy of the second most energetic 
photon ($E_{\gamma 2}$), the recoil mass ($M_{\rm recoil}$) and the polar angle 
of the most forward photon (${\rm max}(|\cos\vartheta_1|,|\cos\vartheta_2|)$), 
for $a_0/\Lambda^2 = -0.1~{\rm GeV}^{-2}$ (Fig.~\ref{a0d}) and 
$a_c/\Lambda^2 = -0.1~{\rm GeV}^{-2}$ (Fig.~\ref{acd}).   
In Figs.~\ref{a0d}-\ref{acd} the predictions obtained by means of the 
complete calculation for the $\nu\bar\nu\gamma\gamma$ final state, including 
$W$-$Z$ interference, (solid histograms), are compared with the results 
relative to the $\nu_{\mu}\bar\nu_{\mu}\gamma\gamma$ process, 
proceeding only via $Z$-boson exchange (dashed histograms). 
For the $\nu\bar\nu\gamma\gamma$ final state, a cut on the recoil mass 
around the $Z$ mass (80~GeV $\leq M_{recoil} \leq 120$~GeV) has been imposed. 
For the sake of comparison, the SM predictions (dotted histograms) normalized 
to the same luminosity are also shown.
The difference
between the solid and dashed histograms indicates a very moderate contribution 
due to $W$-$Z$ interference, even for the large $a_0$ and $a_c$ 
values considered in the simulation.

\begin{figure*}
\begin{center}
\epsfig{file=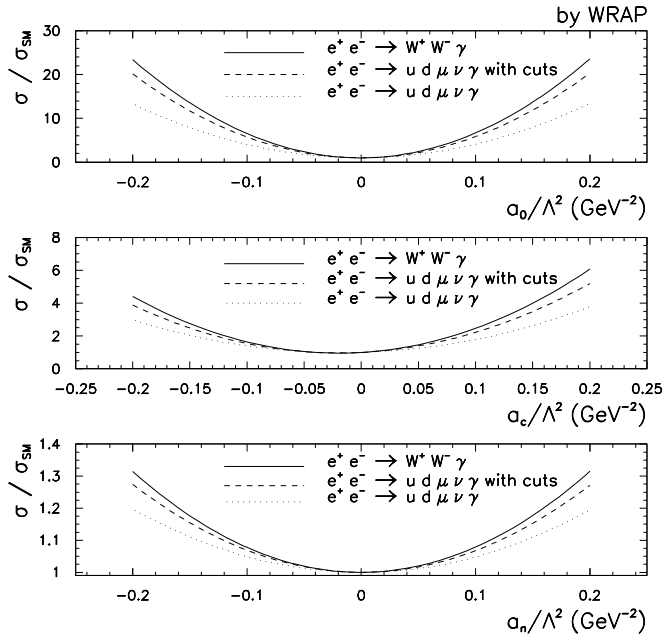,height=10.truecm}
\caption{\small  The sensitivity of the 
$u \bar{d} \mu \bar{\nu}_{\mu}\gamma$ final state to $a_0$, $a_c$ 
and $a_n$ anomalous 
couplings in comparison with the $WW\gamma$ approximation, 
at $\sqrt{s} = $200~GeV.}
\label{a0ac4fg}
\end{center}
\end{figure*}

A similar analysis is shown in Fig. \ref{a0ac4fg}
 for a $4f + \gamma$ final state, as 
computed by means of the exact calculation of {\tt WRAP}, in comparison with the 
$WW\gamma$ approximation considered in the literature~\cite{swer,bela}. 
The sensitivity to the 
anomalous couplings $a_0,a_c,a_n$ is shown for the $4f + \gamma$ final state according to two
different event selections: no cuts on the invariant masses of the decay products 
(dotted line) and cuts on the invariant masses of the decay products around 
the $W$ mass, i.e. 75 GeV $ \leq M_{u \bar{d}, \mu \bar{\nu}_{\mu}} \leq $ 85 GeV, 
(dashed line), in order to disentangle, as much as possible, the 
contributing Feynman graphs 
 with two final-state 
resonant $W$ bosons. The $WW\gamma$ approximation (solid line)    
predicts a quite different sensitivity with respect 
to the complete $4f + \gamma$ calculation.

By considering variations of the anomalous couplings within the allowed 
experimental bounds, differences at the ten per cent level are registered 
between the  $WW\gamma$ approximation and the $4f + \gamma$ prediction, 
when invariant mass cuts are imposed in the $4f + \gamma$ calculation. 
Notice that,  if invariant mass cuts are not considered, 
the $4f + \gamma$ cross section grows up by a factor of two with respect 
to the cross section in the presence of cuts. Therefore, the 
$WW\gamma$ approximation should be employed with the due caution in QAGC studies, 
especially if one takes into account that exact $4f + \gamma$ generators, such as 
{\tt WRAP}~\cite{wrap} and {\tt RacoonWW}~\cite{ddrw}, incorporate the effects of QAGC and 
are at disposal for such experimental studies.

For the sake of simplicity, only numerical results in the 
Born approximation have been presented in the paper. Anyway, as already
remarked and previously discussed in detail in refs.~\cite{nunu,wrap}, 
the effects of ISR can be properly simulated by means of 
{\tt NUNUGPV} and {\tt WRAP} and should be considered 
in the experimental analysis since they tend to diminish
the sensitivity on the QAGC at some per cent level, as 
proved by explicit numerical investigation.

\section{Conclusions}

The search for QAGC in radiative events at LEP demands precise predictions 
for the processes $e^+ e^- \to \nu\bar\nu \gamma\gamma$, $e^+ e^- \to q\bar{q}\gamma\gamma$ and 
$e^+ e^- \to 4$~fermions+$\gamma$. To this end, exact calculations of such 
processes have been presented, incorporating the contribution of QAGC and 
the large effect of ISR. 

By means of the exact calculations, comparisons with approximate results 
existing in the literature have been performed. It turns out that, for 
the $\nu\bar\nu \gamma\gamma$ and $q\bar{q}\gamma\gamma$ final states, 
the $Z\gamma\gamma$ approximation works well, being the interference 
effects present in the complete calculations confined at the per cent level,
if appropriate cuts around the $Z$-boson mass are required. 
It has been also demonstrated that the $\nu \bar\nu \gamma \gamma$ final state 
with appropriate cuts on the recoil mass can be successfully exploited to 
extract limits on neutral QAGC, as a complementary channel to the 
$q \bar{q} \gamma \gamma$ one. As far as 
the 4~fermions+$\gamma$ final states are concerned, significant differences are 
seen between the exact calculation and the $WW\gamma$ approximation, even 
in the presence of invariant mass cuts around the $W$-boson mass.   

In the spirit of the present study, a phenomenological 
analysis of anomalous gauge couplings in radiative events at the 
energies of future linear colliders, including the effects of beam 
polarization and beamsstrahlung, will be given elsewhere.

The exact calculations addressed in the present paper are available in the form of 
improved versions of the event generators {\tt NUNUGPV} and 
{\tt WRAP}~\footnote{\footnotesize The Monte Carlo programs {\tt NUNUGPV} and 
{\tt WRAP} can be downloaded from the WEB site 
{\tt http://decux1.pv.infn.it/\~\,nicrosi/programs.html} }, which can be used for full experimental simulations. 

\newpage\noindent
{\bf Acknowledgements}  \par
We wish to thank 
P.~Bell, D.G.~Charlton, M.~Gataulin, S. Mele, M.~Musy, M.~Thomson, H.~Voss and 
A. Werthenbach for useful information and interest in our 
work.

\end{document}